\documentclass[12pt]{article}
\usepackage{amssymb,amsmath,cite,epsfig}
\usepackage[top=2.5cm, bottom=2.5cm, left=2.5cm, right=2.5cm]{geometry}

\input{tcilatex}
\begin{document}

\begin{titlepage}
\vspace*{3cm}
\begin{center}
{\Large \textsf{\textbf{Second-order corrections to the
non-commutative Klein-Gordon equation with a Coulomb potential}}}
\end{center}
\vskip 5mm
\begin{center}
{\large \textsf{Slimane Zaim$^{a}$, Lamine Khodja$^{b}$ and Yazid
Delenda$^{a}$}}\\
\vskip 5mm
$^{a}$D\'{e}partement de Physique, Facult\'{e} des Sciences,\\
Universit\'{e} Hadj Lakhdar -- Batna, Algeria.\\
$^{b}$D\'{e}partement de Physique, Facult\'{e} des Sciences
Exactes,\\
Universit\'{e} de Béjaia, Algeria.\\
\end{center}
\vskip 2mm
\begin{center}{\large\textsf{\textbf{Abstract}}}\end{center}
\begin{quote}
We improve the previous study of the Klein-Gordon equation in a
non-commutative space-time as applied to the Hydrogen atom to
extract the energy levels, by considering the second-order
corrections in the non-commutativity parameter. Phenomenologically
we show that non-commutativity is the source of lamb shift
corrections.
\end{quote}
\vspace*{2cm}


\noindent\textbf{\sc Pacs numbers}: 11.10.Nx, 81Q05, 03.65.-w
\end{titlepage}

\section{Introduction}

The non-commutative field theory has received a wide appreciation as an
alternative approach to understanding many physical phenomenon such as the
ultraviolet and infrared divergences \cite{1}, unitarity violation \cite{2},
causality \cite{3}, and new physics at very short distances of the
Planck-length order \cite{4}.

The non-commutative field theory is motivated by the natural extension of
the usual quantum mechanical commutation relations between position and
momentum, by imposing further commutation relations between position
coordinates themselves. As in usual quantum mechanics, the non-commutativity
of position coordinates immediately implies a set of uncertainty relations
between position coordinates analogous to the Heisenberg uncertainty
relations between position and momentum; namely: 
\begin{equation}
\left[\hat{x}^\mu,\hat{x}^\nu\right]_{\ast}=i\theta^{\mu\nu},
\end{equation}
where $\hat{x}^{\mu}$ are the coordinate operators and $\theta ^{\mu\nu}$
are the non-commutativity parameters of dimension of area that signify the
smallest area in space that can be probed in principle. This idea is similar
to the physical meaning of the Plank constant in the relation $\left[\hat{x}%
_{i},\hat{p }_{j}\right]=i\hbar \delta_{ij}$, which as is known is the
smallest phase-space in quantum mechanics.

One can study the physical consequences of this theory by making detailed
analytical estimates for measurable physical quantities and compare the
results with experimental data to find an upper bound on the $\theta $
parameter. The most obvious natural phenomena to use in hunting for
non-commutative effects are simple quantum mechanics systems, such as the
hydrogen atom $\left[ 5,6,7\right] $. In the non-commutative space one
expects the degeneracy of the initial spectral line to be lifted, thus one
may say that non-commutativity plays the role of spin.

In this work we present an important contribution to the non-commutative
approach to the hydrogen atom. Our goal is to solve the Klein Gorden
equation for the Coulomb potential in a non-commutative space-time up to
second-order of the non-commutativity parameter using the Seiberg-Witten
maps and the Moyal product. We thus find the non-commutative modification of
the energy levels of the hydrogen atom and we show that the
non-commutativity is the source of lamb shift corrections.

This paper is organized as follows. In section 2 we derive the corresponding
Seiberg-Witten maps up to the second order of $\theta $ for the various
dynamical fields, and we propose an invariant action of the non-commutative
charged scalar field in the presence of an electric field. In section 3,
using the generalised Euler-Lagrange field equation, we derive the deformed
Klein-Gordon (KG) equation. Applying these results to the hydrogen atom, we
solve the deformed KG equation and obtain the non-commutative modification
of the energy levels. In section 4, we introduce the non-relativistic limit
of the NC Klein--Gordon equations and solve them using perturbation theory
and deduce that the non-relativistic NC Klein--Gordon equation is the same
as Schr\"{o}dinger equation on NC space. The last section is devoted to a
discussion.

\section{Seiberg-Witten maps}

Here we look for a mapping $\phi ^{A}\rightarrow \hat{\phi}^{A}$ and $%
\lambda \rightarrow \hat{\lambda}\left( \lambda ,A_{\mu }\right) $, where $%
\phi ^{A}=(A_{\mu },\varphi )$ is a generic field, $A_{\mu }$ and $\varphi $
are the gauge and charged scalar fields respectively (the Greek and Latin
indices denote curved and tangent space-time respectively), and $\lambda $
is the $\mathrm{U}(1)$ gauge Lie-valued infinitesimal transformation
parameter, such that: 
\begin{equation}
\hat{\phi}^{A}\left( A\right) +\hat{\delta}_ {\hat{\lambda}} \hat{\phi}
^{A}\left( A\right) =\hat{\phi}^{A}\left( A+\delta _{\lambda }A\right),
\label{eq:trans}
\end{equation}
where $\delta _{\lambda }$ is the ordinary gauge transformation and $\hat{
\delta}_{\hat{\lambda}}$ is a noncommutative gauge transformation which are
defined by: 
\begin{eqnarray}  \label{eq:tempo1}
\hat{\delta}_{\hat{\lambda}}\hat{\varphi} =i\hat{\lambda}\ast \hat{\varphi}%
,&\qquad&\delta _{\lambda }\varphi =i\lambda \varphi, \\
\hat{\delta}_{\hat{\lambda}}\hat{A}_{\mu } =\partial _{\mu }\hat{\lambda}+i %
\left[ \hat{\lambda},\hat{A}_{\mu }\right] _{\ast },&\qquad&\delta _{\lambda
}A_{\mu }=\partial _{\mu}\lambda.  \label{tempo2}
\end{eqnarray}

In accordance with the general method of gauge theories, in the
non-commutative space, using these transformations one can get at second
order in the non-commutative parameter $\theta ^{\mu \nu }$ (or equivalently 
$\theta $) the following Seiberg--Witten maps \cite{8}: 
\begin{eqnarray}
\hat{\varphi} &=&\varphi +\theta \varphi ^{1}+\theta ^{2}\varphi ^{2}+%
\mathcal{O}\left( \theta ^{3}\right) , \\
\hat{\lambda} &=&\lambda +\theta \lambda ^{1}\left( \lambda ,A_{\mu }\right)
+\theta ^{2}\lambda ^{2}\left( \lambda ,A_{\mu }\right) +\mathcal{O}\left(
\theta ^{3}\right) , \\
\hat{A}_{\xi } &=&A_{\xi }+\theta A_{\xi }^{1}\left( A_{\xi }\right) +\theta
^{2}A_{\xi }^{2}\left( A_{\xi }\right) +\mathcal{O}\left( \theta ^{3}\right)
,  \label{eq:SWM1} \\
\hat{F}_{\mu \xi } &=&F_{\mu \xi }\left( A_{\xi }\right) +\theta F_{\mu \xi
}^{1}\left( A_{\xi }\right) +\theta ^{2}F_{\mu \xi }^{2}\left( A_{\xi
}\right) +\mathcal{O}\left( \theta ^{3}\right) ,  \label{eq:SWM2}
\end{eqnarray}%
where 
\begin{equation}
F_{\mu \nu }=\partial _{\mu }A_{\nu }-\partial _{\nu }A_{\mu }-i\left[
A_{\mu },A_{\nu }\right] .
\end{equation}

To begin, we consider a non-commutative field theory with a charged scalar
particle in the presence of an electrodynamic gauge field in a Minkowski
space-time. We can write the action as: 
\begin{equation}
\mathcal{S}=\int d^{4}x\,\left(\eta^{\mu\nu}\left(\hat{D}_\mu\hat{ \varphi}
\right)^\dagger\ast\hat{D}_\nu\hat{\varphi}+m^{2}\hat{ \varphi}^\dagger\ast 
\hat{\varphi}-\frac{1}{4}\hat{F}_{\mu\nu}\ast \hat{F}^{\mu \nu}\right),
\label{eq:action}
\end{equation}
where the gauge covariant derivative is defined as: $\hat{D}_\mu \hat{%
\varphi }=\left(\partial_\mu+ie\hat{A}_\mu\right)\ast \hat{\varphi}$.

Next we use the generic-field infinitesimal transformations \eqref{eq:tempo1}
and \eqref{tempo2} and the star-product tensor relations to prove that the
action in eq. \eqref{eq:action} is invariant. By varying the scalar density
under the gauge transformation and from the generalised field equation and
the Noether theorem we obtain \cite{9}: 
\begin{equation}
\frac{\partial \mathcal{L}}{\partial \hat{\varphi}}-\partial _{\mu }\frac{%
\partial \mathcal{L}}{\partial \left( \partial _{\mu }\hat{\varphi}\right) }%
+\partial _{\mu }\partial _{\nu }\frac{\partial \mathcal{L}}{\partial \left(
\partial _{\mu }\partial _{\nu }\hat{\varphi}\right) }-\partial _{\mu
}\partial _{\nu }\partial _{\sigma }\frac{\partial \mathcal{L}}{\partial
\left( \partial _{\mu }\partial _{\nu }\partial _{\sigma }\hat{\varphi}%
\right) }+\mathcal{O}\left( \theta ^{3}\right) =0.  \label{eq:field}
\end{equation}

\section{ Non-commutative Klein-Gordon equation}

In this section we study the Klein-Gordon equation for a Coulomb interaction
($-e/r$) in the free non-commutative space. This means that we will deal
with solutions of the U$(1)$ gauge-free non-commutative field equations \cite%
{10}. For this we use the modified field equations in eq. \eqref{eq:field}
and the generic field $\hat{A}_{\mu }$ so that: 
\begin{equation}
\delta \hat{A}_{\mu }=\partial _{\mu }\hat{\lambda}-ie\hat{A}_{\mu }\ast 
\hat{\lambda}+ie\hat{\lambda}\ast \hat{A}_{\mu },
\end{equation}%
and the free non-commutative field equations: 
\begin{equation}
\partial ^{\mu }\hat{F}_{\mu \nu }-ie\left[ \hat{A}^{\mu },\hat{F}_{\mu \nu }%
\right] _{\ast }=0.  \label{eq:freefield}
\end{equation}%
Using the Seiberg-Witten maps \eqref{eq:SWM1}--\eqref{eq:SWM2} and the
choice \eqref{eq:freefield}, we can obtain the following deformed Coulomb
potential \cite{10}: 
\begin{align}
\hat{a}_{0}& =-\frac{e}{r}+\frac{e^{5}}{20\,r^{5}}\left( \theta ^{ij}\right)
^{2}+\mathcal{O}\left( \theta ^{3}\right) , \\
\hat{a}_{i}& =\frac{e^{3}}{4\,r^{4}}\theta ^{ij}x_{j}+\mathcal{O}\left(
\theta ^{3}\right) .
\end{align}%
Using the modified field equations in eq. \eqref{eq:field} and the generic
field $\hat{\varphi}$ so that: 
\begin{equation}
\delta _{\hat{\lambda}}\hat{\varphi}=i\hat{\lambda}\ast \hat{\varphi},
\end{equation}%
the Klein-Gordon equation in a non-commutative space-time in the presence of
the vector potential $\hat{A}_{\mu }$ can be cast into: 
\begin{equation}
\left( \eta ^{\mu \nu }\partial _{\mu }\partial _{\nu }-m^{2}\right) \hat{%
\varphi}\,+\left( ie\eta ^{\mu \nu }\partial _{\mu }\hat{A}_{\nu }-e^{2}\eta
^{\mu \nu }\hat{A}_{\mu }\ast \hat{A}_{\nu }+2ie\eta ^{\mu \nu }\hat{A}_{\mu
}\partial _{\nu }\right) \hat{\varphi}=0.  \label{eq:KGmod}
\end{equation}

Now using the fact that: 
\begin{equation}
\eta ^{\mu \nu }\partial _{\mu }\partial _{\nu }=-\partial _{0}^{2}+\Delta ,
\end{equation}%
and 
\begin{equation}
2ie\eta ^{\mu \nu }\hat{A}_{\mu }\partial _{\nu }=i\frac{2e^{2}}{r}\partial
_{0}-i\frac{2e^{6}}{20\,r^{5}}\left( \theta ^{ij}\right) ^{2}\partial _{0}-%
\frac{e^{4}}{2\,r^{4}}\vec{\theta}.\vec{L},
\end{equation}%
and 
\begin{equation}
-e^{2}\eta ^{\mu \nu }\hat{A}_{\mu }\ast \hat{A}_{\nu }=\frac{e^{4}}{r^{2}}-%
\frac{2e^{8}}{20\,r^{6}}\left( \theta ^{ij}\right) ^{2}-\frac{e^{8}}{%
16\,r^{8}}\left( \theta ^{ij}x_{j}\right) ^{2},
\end{equation}%
where $\vec{L}=r\times p$ and $\vec{\theta}=\left( \theta _{1},\theta
_{2},\theta _{3}\right) .$ setting that $\theta ^{ij}=\epsilon ^{ijk}\theta
_{k}$ ($\epsilon ^{ijk}$ is the Levi-Civita simbol, whish is sub-tensor of
the third level, the lachting symmetry completely ) then the Klein-Gordon
equation \eqref{eq:KGmod} up to $\mathcal{O}\left( \theta ^{3}\right) $
takes the form:

\begin{equation}
\left[ -\partial _{0}^{2}+\Delta -m_{e}^{2}+\frac{e^{4}}{r^{2}}+i\frac{2e^{2}%
}{r}\partial _{0}-\frac{e^{4}}{2\,r^{4}}\vec{\theta}.\vec{L}-\frac{e^{8}}{%
16\,r^{8}}\left( \theta ^{ij}x_{j}\right) ^{2}-i\frac{4e^{6}}{20\,r^{5}}%
\theta ^{2}\partial _{0}-\frac{4e^{8}}{20\,r^{6}}\theta ^{2}\right] \hat{%
\varphi}=0.  \label{eq:temp1}
\end{equation}%
The solution to eq. \eqref{eq:temp1} in spherical polar coordinates $%
(r,\theta ,\phi )$ takes the separable form \cite{11}: 
\begin{equation}
\hat{\varphi}(r,\theta ,\phi ,t)=\frac{1}{r}R(r)Y(\theta ,\phi )\exp
(-\imath Et).
\end{equation}%
Then eq. \eqref{eq:temp1} reduces to the radial equation: 
\begin{multline}
\left[ \frac{d^{2}}{dr^{2}}-\frac{l(l+1)-e^{4}}{r^{2}}+\frac{2Ee^{2}}{r}%
+E^{2}-m_{e}^{2}-\frac{e^{4}}{2\,r^{4}}\vec{\theta}.\vec{L}-\right.
\label{eq:radial} \\
\left. -\frac{e^{6}}{5\,r^{5}}E\theta ^{2}-\frac{e^{8}}{16\,r^{8}}\left(
\theta ^{ij}x_{j}\right) ^{2}-\frac{e^{8}}{5\,r^{6}}\theta ^{2}\right]
R(r)=0.
\end{multline}

In eq. \eqref{eq:radial} the coulomb potential in non-commutative space
appears within the perturbation terms: 
\begin{equation}
-\frac{e^{4}}{2\,r^{4}}\vec{\theta}.\vec{L}-\frac{e^{6}}{5\,r^{5}}E\theta
^{2}-\frac{e^{8}}{16\,r^{8}}\left( \theta ^{ij}x_{j}\right) ^{2}-\frac{e^{8}%
}{5\,r^{6}}\theta ^{2},  \label{eq:perturbation}
\end{equation}%
The first term is obtained in \cite{10} , the second, third and last terms
are new corrections obtained from at the seiberg-witten maps at second
orders ,and induce new splittinges in the $1$S state \cite{10}, which the
non-commutative parameter $\theta $ plays the role of the spin.

\subsection{The solution}

Equation \eqref{eq:radial} has not yet been solved exactly in the presence
of the perturbation terms \eqref{eq:perturbation}, whereas in their absence
its exact solution is available in ref. \cite{13}. To obtain the solution we
choose $\theta =0$ and arrive at: 
\begin{equation}
\left[ \frac{d^{2}}{dr^{2}}-\frac{l(l+1)-e^{4}}{r^{2}}+\frac{2Ee^{2}}{r}%
+E^{2}-m_{e}^{2}\right] R(r)=0.  \label{eq:temp2}
\end{equation}%
This equation is a generalized equation of hypergeometric type $\left[ 14%
\right] $, 
\begin{equation}
R^{\prime \prime }+\frac{\widetilde{\tau }\left( r\right) }{\sigma (r)}%
R^{^{\prime }}+\frac{\widetilde{\sigma }\left( r\right) }{\sigma ^{2}(r)}R=0,
\end{equation}%
where $\sigma (r)=r$, $\widetilde{\tau }=0$, $\widetilde{\sigma }=\left(
Er+e^{2}\right) ^{2}-r^{2}m_{e}^{2}-l(l+1)$. Using the transformation: 
\begin{equation}
R(r)=\phi (r)y(r),
\end{equation}%
equation \eqref{eq:temp2} reduces to an equation of hypergeometric type: 
\begin{equation}
\sigma (r)y^{\prime \prime }+\tau (r)y^{\prime }+\lambda y=0,
\label{eq:hypergeometric}
\end{equation}%
and $\phi (r)$ is the solution of the equation: 
\begin{equation}
\phi ^{\prime }(r)/\phi (r)=\pi (r)/\sigma (r),
\end{equation}%
where the polynomial $\pi (r)$ is defined so that: 
\begin{eqnarray}
\pi (r) &=&\frac{\sigma ^{\prime }-\widetilde{\tau }}{2}\pm \sqrt{\left( 
\frac{\sigma ^{\prime }-\widetilde{\tau }}{2}\right) ^{2}-\widetilde{\sigma }%
+k\sigma }  \notag \\
&=&\frac{1}{2}\pm \sqrt{\left( l+\frac{1}{2}\right)
^{2}-e^{4}-2e^{2}Er+\left( m_{e}^{2}-E^{2}\right) r^{2}+kr}\,.
\end{eqnarray}

According to the Nikiforov-Uvarov (NU) method $[14]$, the expression in the
square--root must be the square of a polynomial, and so one can find new
possible functions for each $k$ as: 
\begin{equation}
\pi (r)=\frac{1}{2}\pm \left\{ 
\begin{array}{l}
\sqrt{m_{e}^{2}-E^{2}}\,r+\sqrt{\left( l+\frac{1}{2}\right) ^{2}-e^{4}},%
\text{ for }k=2e^{2}E+2\sqrt{m_{e}^{2}-E^{2}}\sqrt{\left( l+\frac{1}{2}%
\right) ^{2}-e^{4}} \\ 
\sqrt{m_{e}^{2}-E^{2}}\,r-\sqrt{\left( l+\frac{1}{2}\right) ^{2}-e^{4}},%
\text{ for }k=2e^{2}E-2\sqrt{m_{e}^{2}-E^{2}}\sqrt{\left( l+\frac{1}{2}%
\right) ^{2}-e^{4}}.%
\end{array}%
\right.
\end{equation}%
For all possible forms of $\pi (r)$ we must choose the one for which the
function $\tau (r)=\widetilde{\tau }(r)+2\pi (r)$ has roots on the interval $%
\left. \left[ 0,+\infty \right. \right[ $ and a negative derivative. These
conditions are satisfied by the function: 
\begin{equation}
\tau (r)=1+2\left( \sqrt{\left( l+\frac{1}{2}\right) ^{2}-e^{4}}-\sqrt{%
m_{e}^{2}-E^{2}}\,r\right) ,
\end{equation}%
which corresponds to: 
\begin{eqnarray}
\pi (r) &=&\frac{1}{2}+\sqrt{\left( l+\frac{1}{2}\right) ^{2}-e^{4}}-\sqrt{%
m_{e}^{2}-E^{2}}\,r  \notag \\
\lambda &=&2\left[ e^{2}E-\left( \frac{1}{2}+\sqrt{\left( l+\frac{1}{2}%
\right) ^{2}-e^{4}}\right) \sqrt{m_{e}^{2}-E^{2}}\right] ,
\end{eqnarray}%
and 
\begin{equation}
\phi (r)=r^{\frac{1}{2}+\sqrt{\left( l+\frac{1}{2}\right) ^{2}-e^{4}}}\exp
\left( -\sqrt{m_{e}^{2}-E^{2}}r\right) .
\end{equation}%
The exact energy eigenvalues of the radial part of the Klein-Gordon equation
with a Coulomb potential can be found from the equation: 
\begin{equation}
\lambda +n\tau ^{\prime }+\frac{n\left( n-1\right) }{2}\sigma ^{\prime
\prime }=0,
\end{equation}%
We obtain 
\begin{equation}
E=E_{n,l}^{0}=\frac{m_{e}\left( n+\frac{1}{2}+\sqrt{\left( l+\frac{1}{2}%
\right) ^{2}-\alpha ^{2}}\right) }{\left[ \left( n+\frac{1}{2}\right)
^{2}+\left( l+\frac{1}{2}\right) ^{2}+2\left( n+\frac{1}{2}\right) \sqrt{%
\left( l+\frac{1}{2}\right) ^{2}-\alpha ^{2}}\right] ^{\frac{1}{2}}}%
\,,\qquad \alpha =e^{2}.
\end{equation}

The proper function $y(r)$ in eq. \eqref{eq:hypergeometric} is the
hypergeometric-type function whose polynomial solutions are given by
Rodrigues relation: 
\begin{equation}
y_n(r)=\frac{B_{nl}}{r^{2\sqrt{\left(l+\frac{1}{2}\right)^2-\alpha ^2}}e^{-2%
\sqrt{m_e^2-E^2}\,r}}\frac{d^n}{dr^n}\left( r^{n+2 \sqrt{\left(l+\frac{1}{2}%
\right)^2-\alpha^2}}e^{-2\sqrt{ m_e^2-E^2}\,r}\right),
\end{equation}
where $B_{nl}$ is a normalizing constant. Merging this with the Laguerre
polynomials for $x=2\sqrt{m_e^2-E^2}\,r$, the radial functions are written
as: 
\begin{equation}
R_{nl}(r)=C_{nl}x^{\frac{1}{2}+\sqrt{\left( l+\frac{1}{2}\right)^2 -\alpha^2}%
}e^{-x/2}L_n^{2\sqrt{\left(l+\frac{1}{2}\right)^2-\alpha ^2}}(x),
\end{equation}
where $C_{nl}$ is normalization constant determined by $\int_0^{+%
\infty}R_{nl}^2(r)dr=1$. Thus the corresponding normalized radial functions
are found to be: 
\begin{equation}
R_{nl}(r)=\sqrt{\frac{a}{n+\nu+1}}\left(\frac{n!}{\Gamma\left(
n+2\nu+2\right)}\right)^{1/2}x^{\nu+1}e^{-x/2}L_{n}^{2\nu+1}(x)\,,
\end{equation}
where 
\begin{equation}
\nu=-\frac{1}{2}+\sqrt{\left(l+\frac{1}{2}\right)^2-\alpha^2},
\end{equation}
and $a=\sqrt{m_e^2-E^2}$.

Now, to obtain the modification of the energy levels as a result of the
terms \eqref{eq:perturbation} due to the non-commutativity of space-time, we
use the perturbation theory. To simplify, we take $\theta _{3}=\theta $ and
assume that the other components are all zero, such that $\vec{\theta}.\vec{L%
}=\theta L_{z}$ and $\left( \theta ^{ij}x_{j}\right) ^{2}=\theta ^{2}\left[
\left( r^{2}-z^{2}\right) -2xy\right] $ . In addition we use: 
\begin{equation}
\langle nlm\mid L_{z}\mid nlm^{\prime }\rangle =m_{l}\delta _{mm^{\prime
}}\,\qquad -l\leq m_{l}\leq l,
\end{equation}%
and also the fact that in the first-order perturbation theory the
expectation value of $1/r^{4}$, $1/r^{5}$ and $1/r^{6}$ with respect to the
exact solution of eq. \eqref{eq:temp2}, are given by: 
\begin{eqnarray}
\langle nlm\mid r^{-k}\mid nlm^{\prime }\rangle &=&\int_{0}^{\infty
}R_{nl}^{2}(r)r^{-k}dr\delta _{mm^{\prime }}  \notag
\label{eq:expectationvalue} \\
&=&\frac{2^{k}a^{k}n!}{2\left( n+\nu +1\right) \Gamma \left( n+2\nu
+2\right) }\int_{0}^{\infty }x^{2\nu +2-k}e^{-x}\left[ L_{n}^{2\nu +1}(x)%
\right] ^{2}dx\delta _{mm^{\prime }}  \notag \\
&=&f(k)\qquad k=3,4,5,6.
\end{eqnarray}

We use the relation between the confluent hypergeometric function $F(-n;\nu
+1;x)$ and the associated Laguerre polynomials $L_{n}^{\nu }(x)$, namely: 
\begin{equation}
L_{n}^{\nu }(x)=\frac{\Gamma \left( n+\nu +1\right) }{\Gamma \left(
n+1\right) \Gamma \left( \nu +1\right) }F(-n;\nu +1;x),
\end{equation}%
\begin{multline}
\int_{0}^{\infty }x^{\nu -1}e^{-x}\left[ F(-n;\gamma ;x)\right] ^{2}dx=\frac{%
n!\Gamma (\nu )}{\gamma \left( \gamma +1\right) \cdots \left( \gamma
+n-1\right) }\left\{ 1+\frac{n\left( \gamma -\nu -1\right) \left( \gamma
-\nu \right) }{1^{2}\gamma }\right. + \\
\left. +\frac{n\left( n-1\right) \left( \gamma -\nu -2\right) \left( \gamma
-\nu -1\right) \left( \gamma -\nu \right) \left( \gamma -\nu +1\right) }{%
1^{2}2^{2}\gamma \left( \gamma +1\right) }+\cdots \right. \\
\left. \cdots +\frac{n\left( n-1\right) \cdots 1\left( \gamma -\nu -n\right)
\cdots \left( \gamma -\nu +n-1\right) }{1^{2}2^{2}\cdots n^{2}\gamma \left(
\gamma +1\right) \cdots \left( \gamma +n-1\right) }\right\} .
\end{multline}%
Equation \eqref{eq:expectationvalue} becomes: 
\begin{eqnarray}
\langle nlm\mid r^{-4}\mid nlm^{\prime }\rangle &=&\int_{0}^{\infty
}R_{nl}^{2}(r)r^{-4}dr\delta _{mm^{\prime }}  \notag \\
&=&\frac{16a^{4}n!}{2\left( n+\nu +1\right) \Gamma \left( n+2\nu +2\right) }%
\int_{0}^{\infty }x^{2\nu -1-1}e^{-x}\left[ L_{n}^{2\nu +1}(x)\right]
^{2}dx\delta _{mm^{\prime }}  \notag \\
&=&\frac{8a^{4}n!}{\left( n+\nu +1\right) \Gamma \left( n+2\nu +2\right) }%
\left[ \frac{\Gamma \left( n+2\nu +2\right) }{\Gamma \left( n+1\right)
\Gamma \left( 2\nu +2\right) }\right] ^{2}\times  \notag \\
&&\qquad \qquad \qquad \qquad \times \int_{0}^{\infty }x^{2\nu -2}e^{-x} 
\left[ F(-n;2\nu +2;x)\right] ^{2}dx\delta _{mm^{\prime }}  \notag \\
&=&\frac{4a^{4}}{\left( 2\nu -1\right) \nu \left( 2\nu +1\right) \left(
n+\nu +1\right) }\left[ 1+\frac{3n}{\left( \nu +1\right) }\right.  \notag \\
&&\left. +\frac{3n\left( n-1\right) }{\left( \nu +1\right) \left( 2\nu
+3\right) }\right] \delta _{mm^{\prime }} \\
&=&f(4),
\end{eqnarray}%
\begin{eqnarray}
\langle nlm\mid r^{-5}\mid nlm^{\prime }\rangle &=&\frac{4a^{5}}{\left( 2\nu
-1\right) \left( \nu -1\right) \nu \left( 2\nu +1\right) \left( n+\nu
+1\right) }\left[ 1+\frac{6n}{\left( \nu +1\right) }\right.  \notag \\
&&\left. +\frac{15n\left( n-1\right) }{\left( \nu +1\right) \left( 2\nu
+3\right) }+\frac{5n\left( n-1\right) \left( n-2\right) }{\left( \nu
+1\right) \left( 2\nu +3\right) \left( \nu +2\right) }\right] \delta
_{mm^{\prime }} \\
&=&f(5), \\
\langle nlm\mid r^{-6}\mid nlm^{\prime }\rangle &=&\frac{4a^{5}}{\left( 2\nu
-1\right) \left( \nu -1\right) \nu \left( 2\nu +1\right) \left( n+\nu
+1\right) }\left[ 1+\frac{6n}{\left( \nu +1\right) }\right. \\
&&\left. +\frac{15n\left( n-1\right) }{\left( \nu +1\right) \left( 2\nu
+3\right) }+\frac{5n\left( n-1\right) \left( n-2\right) }{\left( \nu
+1\right) \left( 2\nu +3\right) \left( \nu +2\right) }\right. + \\
&&\left. +\frac{15n\left( n-1\right) }{\left( \nu +1\right) \left( 2\nu
+3\right) }+\frac{5n\left( n-1\right) \left( n-2\right) }{\left( \nu
+1\right) \left( 2\nu +3\right) \left( \nu +2\right) }\right] \delta
_{mm^{\prime }} \\
&=&f(6).
\end{eqnarray}

Putting these results together one gets 
\begin{equation}
\Delta E^{\mathrm{nc}}=-\frac{\alpha ^{2}m_{l}}{2}f(4)\theta -\frac{\alpha
^{3}}{5}\left( E_{n,l}^{0}f(5)+\frac{29}{24}\alpha f(6)\right) \theta ^{2}.
\end{equation}%
The energy shift is due to the terms \eqref{eq:perturbation}. In addition,
the first term of order $\theta $ multiplied by magnetic quantum number
indicates the splitting of states with the same orbital angular momentum
into the corresponding components. This behavior is similar to the Zeeman
effect without spin. The rest of the terms of second order in $\theta $ are
independent of magnetic quantum number, which clearly reflects the existence
of spin. Furthermore it is worth noting that the correction terms containing 
$\theta ^{2}$ are very similar to the spin--spin coupling, thus the
non-commutative parameter $\theta $ plays the role of spin and thus the
degeneracy of levels is completly removed.

The energy levels of the hydrogen atom in the framework of the
non-commutative Klein- Gordon equation are: 
\begin{equation}
\hat{E}=E_{n,l}^{0}-\frac{\alpha ^{2}m_{l}}{2}f(4)\theta -\frac{\alpha ^{3}}{%
5}\left( E_{n,l}^{0}f(5)+\frac{29}{24}\alpha f(6)\right) \theta ^{2}.
\end{equation}

\section{Non-relativistic limit of NC Klein-Gordon equation}

Now we consider the non-relativistic limit of the non-commutative
Klein-Gordon equation \eqref{eq:temp1}. To do this we take the wave function
in the new form: 
\begin{equation}
\hat{\varphi}(r,\theta ,\phi ,t)=\frac{1}{r}R(r)Y(\theta ,\phi )\exp
(-\imath \left( \varepsilon +m_{e}\right) t),
\end{equation}%
where $\varepsilon $ is the non-relativistic energy for which the conditions 
$\left\vert eA_{0}\right\vert \ll m_{e}\left\vert R(r)/r\,Y(\theta ,\phi
)\right\vert $ and $\varepsilon \ll m_{e}$ are valid \cite{15}. Then the
non-relativistic limit of non-commutative Klein-Gordon equation %
\eqref{eq:temp1} is: 
\begin{equation}
\left[ \frac{d^{2}}{dr^{2}}-\frac{l(l+1)}{r^{2}}+\frac{2m_{e}e^{2}}{r}%
+\varepsilon -\frac{e^{4}}{2\,r^{4}}\vec{\theta}.\vec{L}-\frac{e^{8}}{16r^{8}%
}\left( \theta ^{ij}x_{j}\right) ^{2}-\frac{2m_{e}e^{6}}{5r^{5}}\theta ^{2}-%
\frac{e^{8}}{5r^{6}}\theta ^{2}\right] R(r)=0,  \label{eq:new}
\end{equation}%
where the perturbation terms in this equation are the same as those in
equation \eqref{eq:radial} (explicitly given in \eqref{eq:perturbation})
with the replacement $E\rightarrow m_{e}$. Hence in eq. \eqref{eq:new} the
NC terms are similar to the NC Hamiltonian of hyperfine splitting in NC
space \cite{12}. In addition, the new term is similar to the spin-spin
coupling in which the non-commutativity plays the role of the spin. Thus the
energy spectrum and radial wave functions corresponding to $\theta =0$ in
equation \eqref{eq:new} are given by: 
\begin{equation}
\varepsilon _{n}=-\frac{m_{e}\alpha ^{2}}{2\hbar ^{2}n^{2}}
\end{equation}%
and 
\begin{equation}
R_{nl}(r)=\frac{1}{n}\left( \frac{\left( n-l-1\right) !}{a\left( n+l\right) !%
}\right) ^{1/2}x^{l+1}e^{-x/2}L_{n-l-1}^{2l+1}(x),\qquad x=\frac{2}{an}r,
\label{eq:solution0}
\end{equation}%
where $a=\hbar ^{2}/(m_{e}\alpha )$, the Bohr radius of the Hydrogen atom.

Now to obtain the modification of energy levels as a result of the
non-commutative terms in eq. \eqref{eq:new}, we use the first-order
perturbation theory. The expectation value of $r^{-4}$, $r^{-5}$ and $r^{-6}$
with respect to the solution in eq. \eqref{eq:solution0} are given by: 
\begin{eqnarray}
\langle nlm\mid r^{-4}\mid nlm^{\prime }\rangle _{l\succ 0} &=&\frac{4}{%
a^{4}n^{5}l(l+1)(2l-1)(2l+1)(2l+3)}\left[ 3n^{2}-l(l+1)\right] \delta
_{mm^{\prime }},  \notag \\
&=&f(4) \\
\langle nlm\mid r^{-5}\mid nlm^{\prime }\rangle _{l\succ 1} &=&\frac{4}{%
3a^{5}n^{5}(l-1)l(l+1)(l+2)(2l+1)}\times  \notag \\
&&\qquad \qquad \qquad \qquad \times \left[ -1+\frac{5\left(
3n^{2}-l(l+1)\right) }{(2l-1)(2l+3)}\right] \delta _{mm^{\prime }},  \notag
\\
&=&f(5) \\
\langle nlm\mid r^{-6}\mid nlm^{\prime }\rangle _{l\succ 1} &=&\frac{4}{%
a^{6}n^{5}l(l+1)(2l-3)(2l+1)(2l+5)}\times  \notag \\
&&\qquad \times \left[ -\frac{7}{3(l-1)(l+2)}-\frac{3\left(
3n^{2}-l(l+1)\right) }{n^{2}(2l-1)(2l+3)}+\right.  \notag \\
&&\qquad \quad \,\left. +\frac{35\left( 3n^{2}-l(l+1)\right) }{%
3(l-1)(l+2)(2l-1)(2l+1)(2l+3)}\right] \delta _{mm^{\prime }}. \\
&=&f(6)
\end{eqnarray}

\bigskip So the totale modifiction of the energy levels $\Delta E_{nlm}^{%
\mathrm{NC}}$ is given by :

\begin{equation}
\Delta E_{nlm}^{\mathrm{NC}}=-\frac{\alpha ^{2}}{2}m_{l}\theta f(4)-\frac{%
\alpha ^{3}}{5}\theta ^{2}\left( 2m_{e}f(5)+\frac{29}{24}\alpha f(6)\right)
\end{equation}

This energy shift is due to the additional noncommutative terms of eq(%
\eqref{eq:solution0}). The first term is the first-order correction in $%
\theta $ and represents the lamb shift correction for $l\neq 0$ states,
where the energy-level $l$ splits to $2l+1$ sublevels. The second term is
the second order  correction in $\theta $ which represents lamb shift
corrections for $l=0$ states.

This result is very important: as a possible means of introducing electron
spin we replace $l\rightarrow \pm \left( j+\frac{1}{2}\right) $ and $%
n\rightarrow n-j-\frac{1}{2}$, where $j$ is the quantum number associated to
the total angular momentum, then the $l=0$ state have the same total quantum
number $j=\frac{1}{2}$. In this cas the noncomutative value of the energy
levels indicates the splitting of $1s$ states. These results show that the
non-commutative non-relativistic (relativistic) degeneracy is completely
removed. As we have explicitly shown the non-commutativity contributes to
the correction of the lamb shift between these levels. Thus, one can use the
data on the Lamb shift to impose some bounds on the value of the
noncommutativity parameter, $\U{3b8} $. According to \cite{16} the current
theoretical accuracy on the $1s$ Lamb shift is about $14$ kHz. From the
splitting $(63)$, this then gives the bound \ 

\begin{equation*}
\theta \lesssim \left( 2\times 10^{3}\text{GeV}\right) ^{-2}
\end{equation*}%
This is in agreement with other results presented in the ref \cite{3,12} .

\section{Conclusions}

In this work we started from quantum field theory in a canonical
non-commutative space and used the relativistic charged scalar particle in
the Minkowski space-time to find the action which is invariant under the
generalised infinitesimal gauge transformation. By using the Seiberg-Witten
maps and the Moyal product up to second order in the non-commutativity
parameter $\theta$, we generalised the equations of motion and derived the
deformed Klein-Gorden equation for non-commutative Coulomb potential. By
solving the deformed KG equation we found the energy shift up to the second
order of $\theta$, where the first term is proportional to the magnetic
quantum number. This behavior is similar to the Zeeman effect that is
applied in the magnetic field of the system without spin and the second term
is proportional to $\theta^2$, thus we explicitly accounted for a spin
effects in this space. Hence we can say that the Klein-Gordon equation in
non-commutative space at the second order of $\theta$ describes particles
with spin. After that we have obtained the non-relativistic limit of the
deformed Klein-Gorden equation for a non-commutative Coulomb potential. We
showed that the non-relativistic Klein-Gordon equation is the same the Schr%
\"{o}dinger equation for a coulomb potential in non-commutative space. Thus
we came to the conclusion that the non-commutative non-relativistic theory
degeneracy is completely removed and induced the lamb shift where the energy
spectrum of the hydrogen atom depends on the second-order on the
non-commutative parameter $\theta$.

\end{document}